\documentclass{elsart}
\usepackage{graphicx}
\usepackage{amssymb}

\newcommand{\lsim}{\mathrel{\mathop{\kern 0pt \rlap
  {\raise.2ex\hbox{$<$}}}
  \lower.9ex\hbox{\kern-.190em $\sim$}}}
\newcommand{\gsim}{\mathrel{\mathop{\kern 0pt \rlap
  {\raise.2ex\hbox{$>$}}}
  \lower.9ex\hbox{\kern-.190em $\sim$}}}

\begin{document}
\begin{frontmatter}

\title{Diffusion coefficient and acceleration spectrum from direct 
measurements of charged cosmic ray nuclei}
\author[inaf,infn]{Antonella Castellina}
\author[mpi,infn,to]{and Fiorenza Donato}

\address[inaf]{Istituto di Fisica dello Spazio Interplanetario, INAF, Torino - Italy}
\address[mpi]{Max-Planck Institut f\"ur Physik, Munich - Germany}
\address[to]{Dipartimento di Fisica Teorica, Torino - Italy }
\address[infn]{INFN, Sezione di Torino - Italy }

\begin{abstract}
We discuss the potentials of several experimental configurations dedicated to 
direct measurements of charged cosmic ray (CR) nuclei at energies $\gsim$ 100 GeV/n.
Within a two-zone propagation model for stable CRs, we calculate 
light primary and secondary nuclei fluxes for different diffusion and acceleration schemes.
We show that the new detectors exploiting the long and ultra long duration balloon flights
could determine the diffusion coefficient power index $\delta$ through the
measurement of the boron-to-carbon ratio with an uncertainty of about 10-15 \%,
if systematic errors are low enough. 
Only  space-based  or satellite detectors  will be able to determine $\delta$ with very high
accuracy even in the case of important systematic errors, 
thanks to the higher energy reach and the less severe limitations in the exposure.
We show that no uncertainties other than those on $\delta$ affect the determination of the acceleration
slope $\alpha$, so that measures of light primary nuclei, such as the carbon one,
performed with the same experiments, will provide valuable information on the acceleration mechanisms. 
\end{abstract}

\end{frontmatter}

\footnotesize

\section{Introduction} 
Charged particles arriving at Earth with energies between  about $10^7$ eV and
$10^{15}$ eV  are believed to have galactic origin.  Their acceleration is very likely due to the
action of supernova remnants, while their subsequent  diffusion in the Galaxy is driven  by the
turbulent, irregular component of the galactic magnetic field. 
The most abundant galactic cosmic ray particles have energies in the 100 MeV/n - 10
GeV/n range,  where several additional phenomena such as electromagnetic energy losses, convection,
diffusive reacceleration and solar modulation are believed to contribute in shaping their spectra. 
The most realistic description of CR  propagation is given by diffusion models, in which the several
free parameters inherent to a specific model need to be fixed by observations.  
\\
The low-energy tail of the CRs spectrum is shaped by several competing effects, so that it is very 
difficult to disentangle each physical component. On the other hand, most 
data have been collected for energies $\lsim$ 50 GeV/n and their interpretation has not yet led to a
clear understanding of any of the above-mentioned physical ingredients.
The high energy part of the spectrum - let's say $\gsim 10^2$ GeV/n -  
is basically due to 
acceleration and diffusion, all the other effects being minor if not negligible. 
The best-founded  assumptions on the diffusion coefficient and the injection spectrum are
power-laws, and this is what is grossly observed for the flux of nuclei.
\\
A wealth of experimental measurements are available, with different degrees of 
accuracy in the region up to 100 GeV/n; on the contrary,
the higher energy region is poorly known. The most recent direct measurements in this region 
have been provided by  the series of balloon flights of JACEE \cite{jac} and RUNJOB 
\cite{runj}, while most of the data results from indirect measurements, by means of 
ground arrays observing air showers.
\\
New data have recently been added by the ATIC Collaboration \cite{Atic}, 
connecting the lowest energy region to the highest energy available data. The new 
CREAM project \cite{CreamProp} has been developed and is now in data taking 
phase, with the aim of dramatically increasing the available
statistics in the energy region up to 500 TeV and possibly more, 
thus making new and precise measurements of the spectral characteristics of CRs.
\\
In the present paper, we explore the performances required for new detectors 
to disentangle the fundamental parameters describing the propagation of nuclei 
at energies above about 100 GeV/n and below the knee region. 
In Sect. \ref{Exp} we report on the status of  experimental projects aimed at the 
measurement of CR nuclei at energies $\gsim$ 1 GeV/n. In Sect. \ref{prop} 
we describe a diffusion model for galactic CRs, and highlight 
the main features affecting the propagation of particles in the higher energetic range.
Sect. \ref{sim} summarizes the method employed for the simulation of the experimental conditions.
In Sect. \ref{parliamone} we present our results focusing on the possibilities to measure
the diffusion coefficient slope by means of the boron-to-carbon (B/C) data and the acceleration
spectrum by means of primary light nuclei fluxes such as the carbon one. In Sect. 
\ref{concludiamo} we draw our conclusions.

\section{Experimental status and projects}
\label{Exp}
\begin{table}[htb]
\caption{{\it Balloon and space instruments and new projects. Experiments covering an energy 
range below 1 GeV/n are not considered in the Table. For all the new projects on satellite we assume 
1000 days of data recording. At least 10 events are expected to be detected
by ATIC,CREAM and ACCESS above the maximum quoted energy.}}
\label{ta:esper}
\newcommand{\m}{\hphantom{$-$}}
\newcommand{\cc}[1]{\multicolumn{1}{c}{#1}}
\renewcommand{\tabcolsep}{0.3pc} 
\renewcommand{\arraystretch}{1.2} 
\begin{tabular}{|@{}c|@{}c|@{}c|@{}c|@{}c|@{}c|}
\hline
Experiment & Years & Elements & Energy & $\Gamma$ & Exposure \\
                 &  &          & GeV & m$^{2}$ sr & (m$^{2}$ sr days) \\ \hline \hline
Balloon \ instruments& & & & & \\ \hline
{\sc mubee} \ \cite{Zat} & 1975-87 & 1 $\leq$ Z $\leq$ 26 & 10$^{4}$-3 $\cdot$ 10$^{5}$ & 0.6 & 22 \\
{\sc sanriku} \ \cite{sanr} & 1987-91 & 8 $\leq$ Z $\leq$ 26 & 2 $\cdot$10$^{2}$-3$\cdot$ 10$^{4}$ & 4.81 & 4.45 \\
{\sc jacee} \ \cite{jac} & 1979-95 & 1 $\leq$ Z $\leq$ 26 & 2 $\cdot$10$^{3}$-8$\cdot$ 10$^{5}$ & 2-5 & 65-107 \\
{\sc runjob} \ \cite{runj} & 1995-99 & 1 $\leq$ Z $\leq$ 26 & 10$^{4}$-3 $\cdot$10$^{5}$ & 1.6 & 43  \\
{\sc tracer} \ \cite{Tra} & 2002-04 & 5 $\leq$ Z $\leq$ 28 & 1.6 $\cdot$10$^{2}$-1.6 $\cdot$10$^{4}$  \footnotemark & 5 & 70 \\ 
{\sc atic} \ \cite{Atic}  & 2000-03 & 1 $\leq$ Z $\leq$ 28 & 10-10$^{5}$ & 0.25 & 3.5 \\ 
{\sc cream} \ \cite{CreamProp} & 2004$\rightarrow$ & 1 $\leq$ Z $\leq$ 28 & 10$^{3}$-10$^{6}$ & 0.5/1.3 \footnotemark & 35-140\footnotemark \\ 
\hline \hline
Space \ instruments & & & & & \\ \hline
{\sc proton}\ 1-4 \cite{Gri}  & 1965-68  & All particles,p,He         & 100-10$^{6}$ & 0.05-10   & 5-2000 \\
{\sc sokol} \ \cite{Sok}	 & 1984-86  & 1 $\leq$ Z $\leq$ 26 & 2 $\cdot$10$^{3}$-10$^{5}$ & 0.026	 & 0.4 \\
{\sc heao-3} \ \cite{Heao}	 & 1979-80  & 4 $\leq$ Z $\leq$ 28 & 9.6-560   \footnotemark[1] & 0.14	& 33 \\
{\sc crn} \ \cite{CRN,Mul2}	 & 1985     & 4 $\leq$ Z $\leq$ 26 & 7$\cdot$ 10$^{2}$-3 $\cdot$10$^{4}$ & 0.1-0.5/0.5-0.9\footnotemark[2]  & 0.3-3 \\
 \hline \hline
Projects  & & & & & \\ \hline
{\sc nucleon-klem} \ \cite{Klem} & & 1 $\leq$ Z $\leq$ 26 & 10$^{2}$-10$^{7}$ & 0.19 & 190\\ 
{\sc ams-02} \ \cite{ams} & &  1 $\leq$ Z $\leq$ 26& 1-10$^{4}$ & 0.5 & 500  \\
{\sc access} \ \cite{Acc} & & 1 $\leq$ Z $\leq$ 28 & 10$^{3}$-5 $\cdot$10$^{6}$ & 1/8 \footnotemark[2] & 300-8000 \\ 
{\sc proton 5} \ \cite{pr5} & & 1 $\leq$ Z $\leq$ 28 & 10$^{3}$-10$^{7}$ & 18& 18000 \\ 
{\sc inca-cstrd} \ \cite{Inc}& & 1 $\leq$ Z $\leq$ 28 & 10$^{5}$-10$^{7}$ & 48 & 48000 \\ 
\hline \hline
\end{tabular}\\[2pt]
{\small
(1) the energy range refers to oxygen;\\
(2) the two numbers refer to low and high Z respectively;\\
(3) CREAM had a succesful 40 days flight in December 2004-January 2005, but it is planned 
to exploit ULDB flights of $\simeq$ 100 days in the near future.}
\end{table}
In Table \ref{ta:esper}, we collected the past and current experiments as well as new projects 
directly measuring charged CRs nuclei in the energy range above 1 GeV/n and spanning a wide range in Z.
\\
Direct measurements of charge and energy of CR nuclei started in the 
sixties, exploiting the balloon and spacecraft techniques. Scintillators
and Cerenkov detectors were employed for charge measurement, while ionization 
calorimeters measured the nucleus energy \cite{Dwy,Lez,Sim,Ort,Mah,Gri,Heao}.
\\ 
Four experiments \cite{jac,runj,Zat,sanr} exploited passive detectors like nuclear 
emulsions and X-ray films: due to their sensitivity also to inclined tracks, 
they allowed to reach a much bigger exposure compared to active 
detectors, thus in principle exploring the highest energy region (about 10 to 1000 TeV).
 The most precise measurements on relative abundances of elements with  Z from 4 up to 28 
came from the HEAO-3 mission, whose data have been extensively used to tune and check models. 
AMS-01 (a spectrometer flown  on the space shuttle Discovery including a permanent magnet, 
a tracker, time-of flight hodoscopes and a Cerenkov counter) brought 
further information on proton and helium up to about 100 GeV/n \cite{Ams1}. \\
Among the present experiments, a fully active bismuth germanate (BGO) 
calorimeter flew on ATIC \cite{Atic}, while 
TRACER \cite{Tra} employed a transition radiation detector.\\ 
The development of the Long Duration technology is now giving the
possibility to balloon experiments to fly for more than 20-30 days, thus 
reaching much higher exposures. 
CREAM \cite{CreamProp} employs for the first time  both a calorimeter and a transition 
radiation detector, thus allowing an intercalibration of the energy measurement with two
independent techniques with different systematic biases. This apparatus is planned to 
exploit the Ultra Long Duration Balloon flights technology, which relies on  new superpressure 
closed balloons with a pumpkin-shape design and is
now in test phase; it will hopefully fly up to 100 days. 
\\
New space instruments have also been proposed: a) the NUCLEON russian satellite program which will
include the KLEM detector (silicon microstrips and scintillator strips) \cite{Klem}; b) a ionization 
calorimeter on a series of Proton-5 satellite missions \cite{pr5}, 
with the aim of measuring the spectrum of elements up to 
more than 1000 TeV, reaching an accuracy of 2-5\% in the slope determination;
c) INCA \cite{Inc}, a ionization neutron calorimeter with 48 m$^2$sr acceptance; its main aim being
the study of the electron spectrum, it is expected to measure also primary nuclei in the knee
region. A proposal for an integration of this calorimeter with a Compton 
scatter transition radiation detector has been recently presented \cite{intr}.
The ACCESS \cite{Acc} project was not chosen for the last MIDEX phase by NASA, but has been 
included in the hope of a reproposition as a free-flyer.
\\
We should also mention the sophisticated spectrometers which were flown on 
balloons, with the main aim of measuring antimatter and light isotopes \cite{lowE}.
They are not included in the Table because of their quite
different aim, but they also gave very precise information on the proton 
and helium flux up to around 100 GeV/n; the energy range has been extended by the  Bess-TeV \cite{best} 
upgrade of Bess-98, reaching about 500 GeV  and 200 GeV/n for proton and helium nuclei respectively.
The new AMS-02 \cite{ams} large acceptance magnetic spectrometer has been conceived 
to study origin and structure of the dark matter and to measure antinuclei; it will also
be able to extend the knowledge on the composition of charged CRs from the 100 MeV/n region
up to about 1 TeV/n.

\section{Propagation of CRs in the Galaxy} 
\label{prop}
Despite scarce theoretical and observational knowledge of the key parameters
responsible for acceleration and propagation of galactic CRs, phenomenological models able to reproduce
data  can be built.  The most realistic propagation models are the diffusion
ones, even if the so--called leaky box model has been often preferred in the past for its simplicity
\cite{berezinsky}. 
In Refs. \cite{usineI,usineIbis}, a two-zones diffusion model has been developed and shown to 
reproduce several observed species in the low-energy part of the
CR spectrum ($\sim$ 0.1-100 GeV/n). In this model, the Galaxy has
been cylindrically shaped (with $R$=20 kpc), with a
thin disc (half-hight $h$=0.1 kpc) containing the sources and the interstellar medium (ISM)
 surrounded by  a diffusive halo of  half-thickness $L\sim$ 2-15 kpc. 
The transport equation  for the nucleus $j$ in a diffusion model is in principle valid 
in a very wide range of energies and can be written as:
\begin{eqnarray}
     K(E)\left(\frac{\partial^{2}}{\partial z^{2}}+
          \frac{1}{r}\frac{\partial}{\partial r}
        (r\frac{\partial}{\partial r})\right)N^{j}(E,r,z) -V_{c} \frac{\partial}{\partial z}
      N^{j}(E,r,z)
      \\ \nonumber
      +2h\delta(z) \left(
         q^j_0Q_j(E)q(r)+\sum_{k=1}^{j-1}\Gamma^{kj}N^{k}(E,r,0)
         -\Gamma^j N^{j}(E,r,0)\right)\nonumber\\
        =2h\delta(z) \frac{\partial}{\partial E}
        \left\{ b^j(E)N^j(E,r,0) - d^j(E)\frac{\partial}{\partial E}
       N^j(E,r,0)\right\}\nonumber
       \label{eq:diff_eq}
\end{eqnarray}    
Steady-state has been assumed and  $N^j(E,r,z)$ is the differential density of the nucleus $j$ as
a function of energy $E$ and galactic coordinates ($r,z$).  In this equation, the first term
represents spatial diffusion with  diffusion coefficient $K(E)$, which has been assumed to be
independent of Galactic coordinates. $V_c$ is the convection
velocity, assumed here to be constant throughout the Galaxy (except in the thin disk) and to be
directed outward along the z-direction.  The term
proportional to $2h$ in the left-hand side of Eq. (1) takes into account all the sources of cosmic
rays: primary sources with injection spectrum $Q^j(E)$, secondary spallative production from
heavier nuclei and destructive reactions (radioactive-decay terms have been omitted for
simplicity). The right-hand side of Eq. (1) contains -- through the coefficients 
$b^j(E)$ and $d^j(E)$ -- the terms  responsible of the energetic
changes suffered by charged particles during propagation:  Coulomb, ionization and adiabatic
expansion losses, and gains due to second order reacceleration.  Diffusive, or continuous, 
reacceleration is due to the scattering of charged particles on the magnetic turbulence in the
interstellar hydrodynamical plasma. The diffusive reacceleration coefficient is related to the
velocity of  such disturbances, called the Alfven velocity $V_A$, and is naturally connected to
the space diffusion coefficient $K(E)$.  For example, diffusive reacceleration contributes
significantly in shaping the boron--to--carbon (B/C) ratio at kinetic energy per nucleon E around
1 GeV/n.  Indeed, all energy losses and gains are effective only in the
low-energy tail of the CR spectrum and thus are irrelevant in the analysis carried in
the following of our paper, which deals with E$\gsim$ 100 GeV/n.
The so-called sporadic, or distributed, reacceleration has been
considered in the literature  \cite{wandel,berezhko} and is based on the possibility that
CRs be reaccelerated during their wandering in the Galaxy by supernova remnants (SNRs) and gain a
small amount of energy at each encounter. 
\\
Convection may dominate at low energies, depending on the value of $V_c$  and if $\delta$ is large.
It may compete with diffusion up to  few tens of GeV/n, 
but its role becomes negligible at higher energies. The importance of 
spallation processes 
depends on the nucleus, but at energies $\gsim$ 100 GeV/n they are not much relevant 
except for heavy nuclei (e.g. iron) \cite{david_HE}.   We also note that
the different assumptions on the distribution of the ISM - homogeneously distributed (as assumed in
our analysis) or spatial dependent - are irrelevant at high energies and for the species
considered in our analysis. 
\\
Diffusion depends on the rigidity $R$ ($R=p/Z$) of the particle and the diffusion coefficient is
usually assumed to have the form: $K(E)=K_0 \beta R^\delta$.  $K_0$ is linked to the
level of the hydromagnetic turbulence and $\delta$ to the density spectrum of these irregularities
at different wavelength.  The Kolmogorov theory for the turbulence spectrum predicts
$\delta$=1/3, while  the case for a hydromagnetic spectrum with $\delta$=1/2  has been obtained by
Kraichnan \cite{kraichnan}. On the other side, the phenomenological  interpretations of CR
spectra have not led to fix $\delta$ because of the
complicated treatment of the physical phenomena  which are relevant at the energies in which  most 
cosmic particles have been collected. Fits to the B/C ratio within the diffusion model of Eq. (1)
prefer high values for $\delta$ ($\sim$ 0.5-0.75), while the Kolmogorov spectrum turns out to be 
disfavored \cite{usineI,usineIbis,jones}. The $\delta$=1/3 case can
reproduce quite well the peculiar B/C peak  observed at $\sim$ 1 GeV/n but it tends to
overestimate the data at increasing energies,   predicting a quite flat B/C ratio (see
also later).   The highest values of $\delta$ seem on the other side not very realistic. 
In fact they are only marginally consistent with observations of interstellar scintillation
\cite{armstrong,turbo} and are highly incompatible with the level of anisotropy
measured at 1-100 TeV \cite{anis}. 
Data on CR fluxes  do not extend to sufficiently high energies 
and are not sufficiently precise in order to constrain  $\delta$.
\\
The acceleration spectrum of primary CRs is believed to be determined by supernovae (SN)  remnants and
super-bubbles \cite{drury}, which are the only known engines in the Galaxy able to provide the right
amount of energy to particles up to $\sim  Z \times 10^{14}$ eV. The acceleration spectrum  follows  a
power-law in momentum, $Q(E)\propto p^{-\alpha}$, with  $\alpha$  located somehow between 2.0 and 2.5.
Even if a precise value for $\alpha$ cannot be predicted, the effective spectral index as derived by 
several indirect observational tests and by numerical studies is close to $\alpha \sim$ 2.0-2.1
\cite{drury} (and refs. therein), \cite{ber94}.  Very recently, the TeV $\gamma$-ray image of a SN
remnant -  already observed in the X-ray spectrum with a very similar morphology -  strongly indicates
that high energy nuclei are accelerated in this site \cite{hess}.  The $\gamma$-ray data are well
reproduced by a photon spectral index which seems to prefer low $\alpha$ values, even if the
experimental  sensitivity is not yet sufficient to put strong constraints.  Higher values of $\alpha
\sim$  2.4 are favored by propagation  models for GCRs with very low $\delta$, such as the Kolmogorov
spectrum. 
\\
The above-described diffusion model has been tested on the B/C and sub-Fe/Fe ratios and shown to fit well 
existing data, which we remind lie in the low-energy tail of the galactic CR spectrum 
\cite{usineI,usineIbis}.
The propagation parameters able to reproduce such ratios  give also rise to a secondary antiproton flux which is
in very good agreement with measurements \cite{pbar_sec}. A further, yet weaker, degree of consistency is given
by an analysis of radioactive isotopes \cite{beta_rad}. 
In this case, the diffusion model has been modified 
in order to take into account a local under-dense region, relevant when dealing with short-living species.
The model has also been validated at higher energies through a study of the mean logarithmic mass of the CR 
\cite{david_HE} beam.
For further details on this  propagation model and considerations on 
other possible approaches we refer to Ref. \cite{revue} and references therein.

\section{Detector design and simulation}
\label{sim}

To pursue the objective of extracting information about crucial propagation 
parameters in the high energy region, an experimental apparatus should have the 
following characteristics:
\\
(i) A large exposure, to have enough statistics: to detect about 10 nuclei 
at energies above 10 TeV/n, if taking the elemental fluxes from \cite{wsoo}, 
a collecting power $\Gamma \simeq$53 m$^2$sr days is to be reached  for carbon nuclei, 
while more than 250 m$^2$sr days are necessary for Iron.
\\
(ii) A good energy resolution: this is especially crucial to detect changes of slope
in the energy spectra (e.g. the knee in the proton spectrum) or to study spectral
smoothness. A constant 40\% energy resolution is enough to see an increase in the
spectral slope of about 0.3 crossing the knee. The same resolution or better is required
to find small deviations from smoothness, of the order of 10\% \cite{pr5}.
\\
(iii) A charge resolution such as to distinguish e.g. carbon from boron nuclei;
models ask for a B/C ratio of some percent around 1 TeV/n, so that a
resolution of about 0.2 charge units is needed.
\\
Combining different detectors in the same experiment gives a powerful tool to overcome
individual technical limitations: for example, redundant measurements of energy allow
a cross-calibration of the detectors, thus overcoming the problem of direct energy calibration
which at the highest energies explored is possible for TRDs but not for calorimeters.
\\
A very careful study of systematic errors is mandatory, as these are the main cause of uncertainty 
in all the measurements. Exposures and efficiencies in selecting and
tracking the events can be studied by means of Monte Carlo simulations; for both energy and 
charge determination, redundancy can help in reducing the ambiguities.
Although a serious discussion of systematics depends on each specific detector, a constant 
contribution to the overall error on the flux measurements will be introduced
in our analysis. 
\\
In order to explore the performances of new detectors, a simulation has been built 
with the following ingredients:\\
(a) the input flux; the particle spectrum is numerically given by the propagation model.
Different power laws with increasing slope in contiguous energy ranges are fitted to the
spectrum to better follow the expected behavior.
The flux is normalized to the best experimental low energy datum (e.g. the flux as derived by 
\cite{Heao} at 10.6 GeV/n).\\
(b) the experimental apparatus, in terms of collecting power and energy
resolution. We use $\Gamma$ = 1.3, 5 and 10 m$^{2}$sr, similar to the quoted geometrical 
factors of some of the current and future experiments and a set of exposure
times of 30, 100 and 1000 days, which roughly correspond to long and ultra
long duration balloon flights and satellite conditions, respectively. 
\\
The expected number of events above a given energy E for an assumed
collecting power $\Gamma$ and input spectrum $dN/dE = C_{0} E^{-\gamma}$ m$^{-2}$s$^{-1}$sr$^{-1}(GeV/n)^{-1}$ 
can be written as:
\begin{equation}
N(>E) = \Gamma C_{1} E^{-\gamma+1}
\label{eq:nev}
\end{equation}
where $C_{1}$ depends on the slope and normalization of the input flux. 
\\
Following the approach of \cite{how}, the cumulative distribution function for E in the considered energy range
can be written as
\begin{equation}
\Phi(E) = 1 - \frac{N(>E)-N(>E_{b})}{N(>E_{a})-N(>E_{b})} \ \ \ [E_{a} \leq E \leq E_{b}]
\end{equation}

and the probability distribution function for the events is obtained by
differentiation 
\begin{equation}                                               
d\Phi(E)/dE =\frac{(\gamma-1)}{E_{a}^{-\gamma+1}-E_{b}^{-\gamma+1}} E^{-\gamma}
\end{equation}

The number of events N of Eq.\ref{eq:nev} is the mean value of the Poisson distribution of the true number of events 
in a given range  $[E_{a}-E_{b}]$; the poissonian fluctuation is computed in small energy bins (equal bins in logarithmic
scale so that, for any interval {\it j}, $E_{j+1}/E_{j}$=constant) covering the full interval,
in order to correctly weight also the bins with few events.
The energy of each {\it i}-th event is then randomly sampled in the energy
interval $[E_{a}-E_{b}]$ from the power law spectrum; being $r_i$ a standard uniform variate
\begin{equation}
\Phi(E_{i}) = \int_{E_{a}}^{E_{i}}d\Phi(E')/dE' dE' = r_{i}
\end{equation}
and
\begin{equation}
E_{i} = \{ r_{i} (E_{b}^{-\gamma+1}-E_{a}^{-\gamma+1})+E_{a}^{-\gamma+1} \}^{\frac{1}{1-\gamma}}
\end{equation}
In each interval [E$_{j}$-E$_{j+1}$], the mean energy is 
\begin{equation}
<E> = (\frac{\gamma}{\gamma-1})(\frac{E_{j}^{-\gamma+1}-E_{j+1}^{-\gamma+1}}{E_{j}^{-\gamma}-E_{j+1}^{-\gamma}})
\end{equation}
The detector response is modeled  as a Gaussian distribution with mean equal to the input 
energy and width given by the energy resolution. In the following, a constant 40\% energy resolution will be
assumed; the case for resolutions either decreasing or increasing with energy will however be checked.
The detection efficiency is here assumed not to depend on energy and will be included in the systematic
uncertainties.

\section{Results and discussion}
\label{parliamone}
The aim of this section is to study the potentials of different experimental configurations 
in connection with the determination of the diffusion coefficient slope $\delta$ and the 
acceleration power spectrum $\alpha$. We show how well their physical value could be singled out
with high-energy data on the B/C ratio and the carbon flux.

\subsection{B/C and the diffusion coefficient slope}
\begin{figure}
\includegraphics[width=0.99\textwidth]{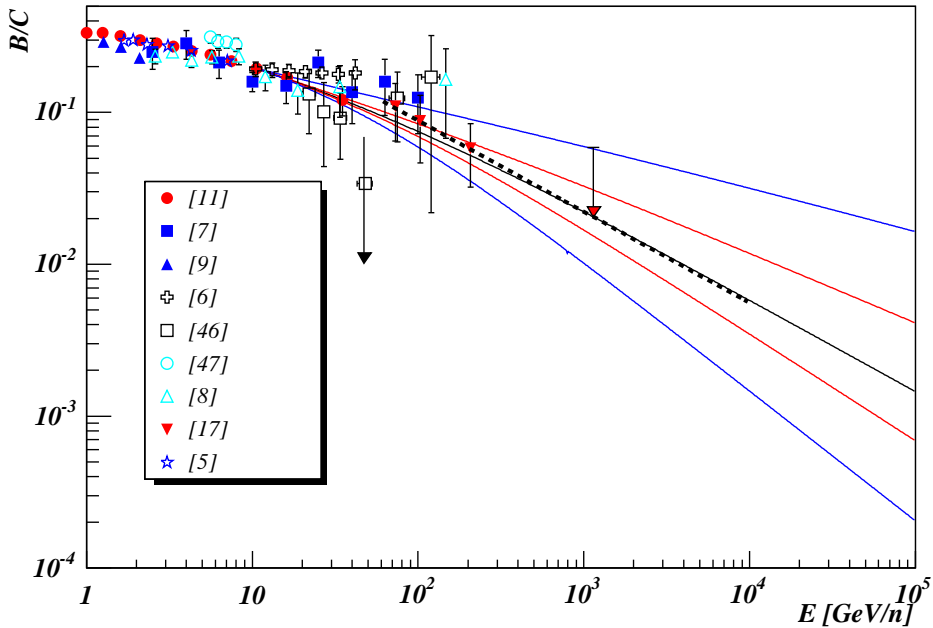}
\caption{\it{B/C ratio as a function of kinetic energy per nucleon. Solid curves:
theoretical predictions corresponding, from top to bottom, 
to cases 1-2-3-4-5, respectively (see text for details).
Dashed line: simple power law spectrum B/C=1.4 E$^{-0.60}$.}}
\label{bc_theo_data}
\end{figure}
\begin{figure}[htp]
\centering
\includegraphics*[width=0.75\textwidth]{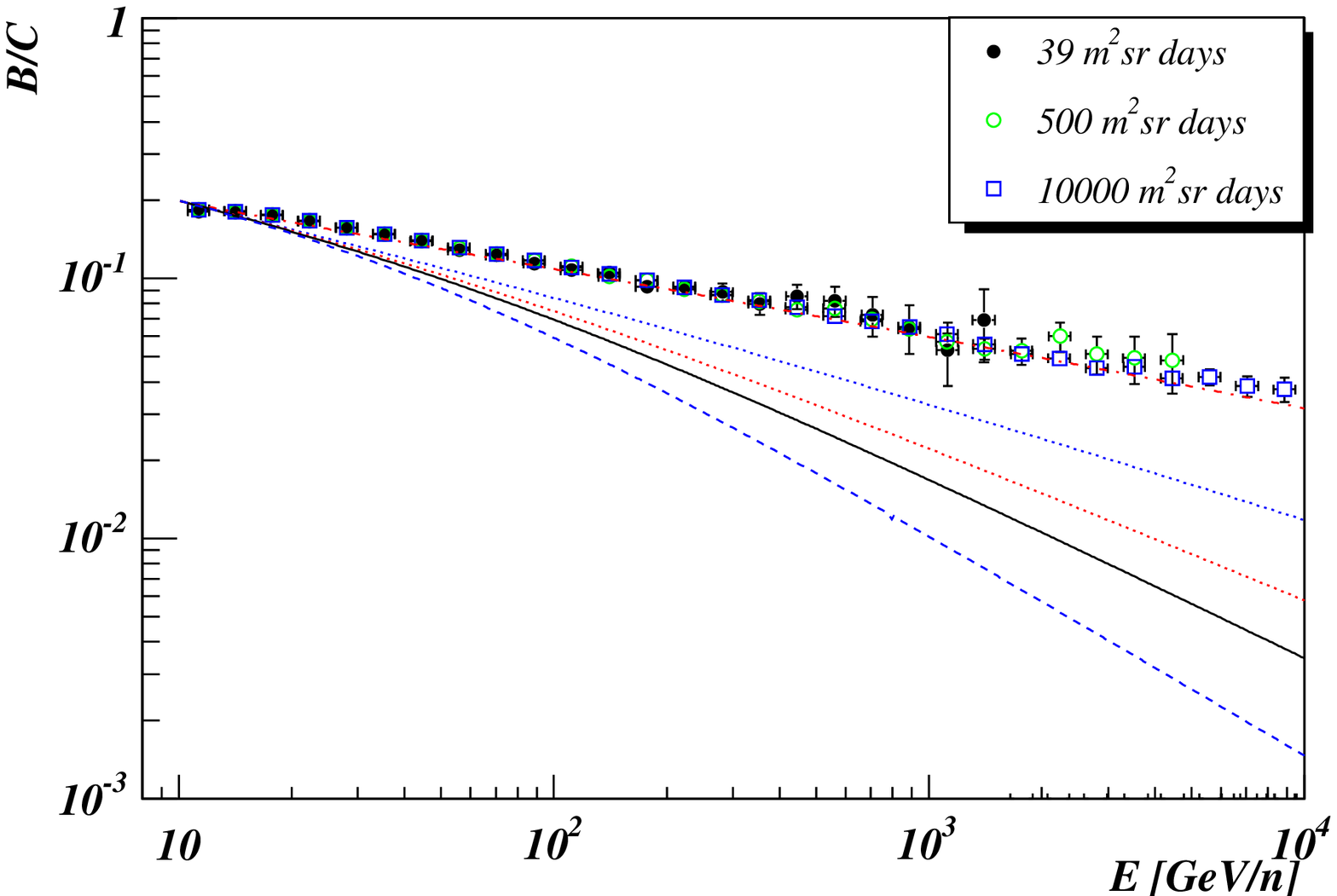}
\includegraphics*[width=0.75\textwidth]{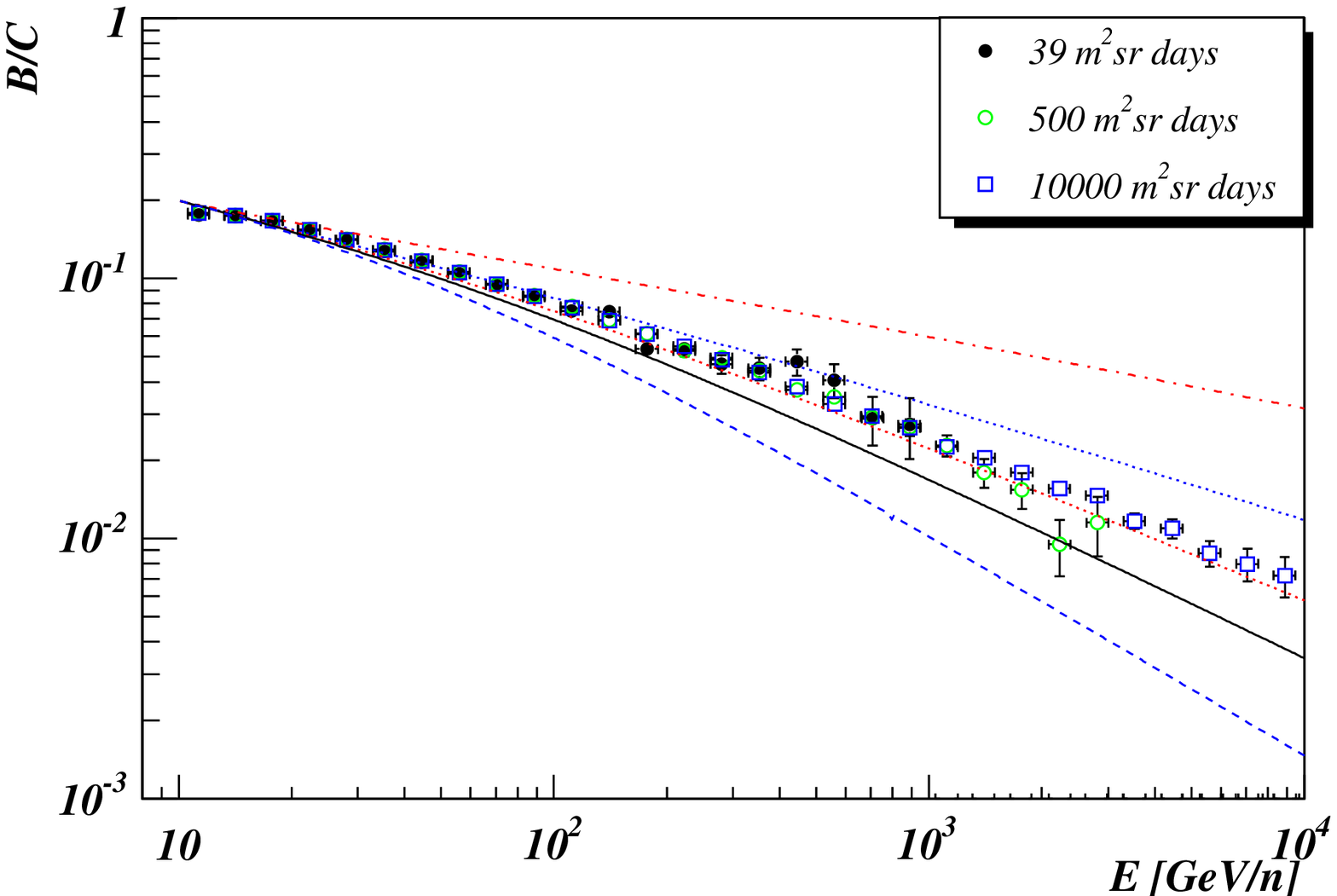}
\hfill
\caption{\it{B/C ratio as expected from input case 1 (upper panel) and 3 (lower panel) for 
different experimental exposures.
The expected ratios from cases 1 to 5 (from top to bottom, respectively) are also shown.}}
\label{fi:BCattb}
\end{figure}
\begin{figure*}[!htbp]
 \includegraphics[width=.99\textwidth]{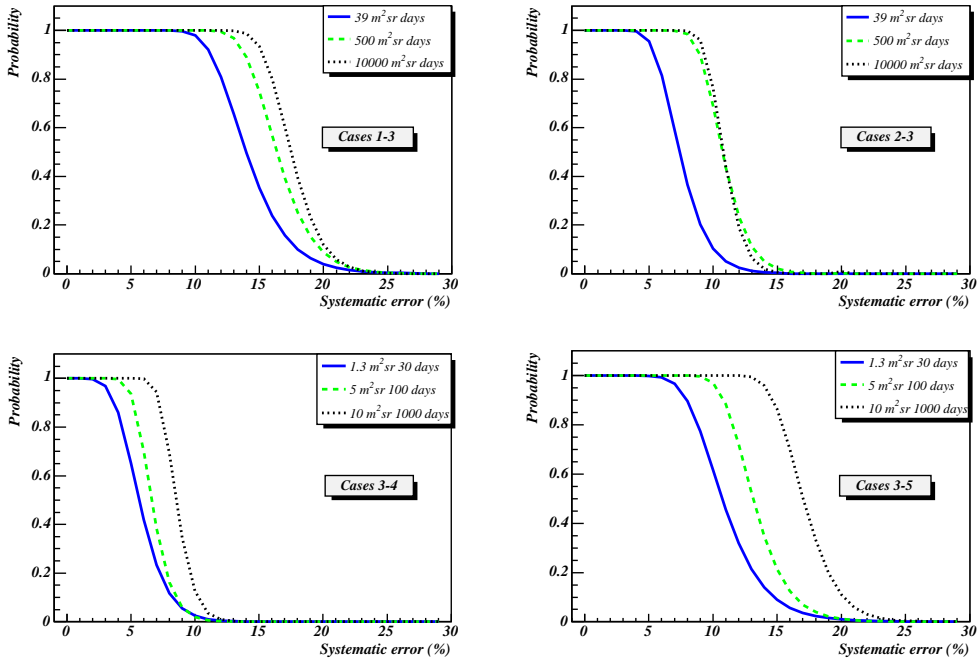}
 \caption{\it{Probability of distinguishing between two input models vs systematic error for three 
different collecting powers. Remind that, for cases 1-5, $\delta$=0.30,0.46,0.60,0.70,0.85 respectively.}}
 \label{fi:probGF}
\end{figure*}

The boron-to-carbon ratio has always been considered the best quantity
to study diffusion properties. At energies around GeV/n, this ratio is strongly 
affected  by the low-energy phenomena described in Sect. \ref{prop}. In particular, 
the observed bump at $\sim$ 1 GeV/n can be naturally explained by diffusive reacceleration, 
without invoking any artificial break in the diffusion coefficient. \\
The most precise B/C data
have been obtained in the energy range 0.6-35 GeV/n by the HEAO-3 experiment \cite{Heao},
the highest energy points having been collected by CRN \cite{CRN}. 
In Fig. \ref{bc_theo_data}  we plot a collection 
of data on the B/C ratio giving emphasis to the higher energy part of the spectrum. 
Along with the experimental data, we also plot the theoretical predictions calculated with 
the full diffusion model briefly outlined in Sect. \ref{prop}.
\\
We have chosen five illustrative configurations of the propagation parameter space, 
which will be also employed in the following of our analysis as bench marks. 
At the energies we deal with, the parameter which mostly shapes the B/C ratio is the 
diffusion coefficient slope, namely $\delta$. The constant $K_0$ in the diffusion coefficient enters with 
the source abundances in the global normalisation of B/C, so that 
its precise value is of scarce relevance in the B/C predictions.
We thus identify cases 1-2-3-4-5 with 
$\delta=0.3, 0.46, 0.6, 0.7, 0.85$, respectively.
All the other propagation parameters ($K_0, L, V_c, V_A, \alpha$)
have been fixed according to the best fits to B/C data \cite{usineIbis}. Indeed 
their exact value is not very important when 
working with a secondary-to-primary flux ratio at energies $\gsim$ 100 GeV/n.
The dashed line in Fig.\ref{bc_theo_data} corresponds to the power law
$1.4 E^{-0.60}$, where the coefficient 1.4 has been fixed to the theoretical 
prediction for the $\delta$=0.6 case at 1000 GeV/n. This line shows that 
for energies $\gsim$ 1000 GeV/n the B/C ratio can be well approximated by a 
power law in energy with index $-\delta$, as one expects from the high energy 
limit: secondary/primary $\propto K(E)^{-1} \propto R^{-\delta}$. 
At E=100 GeV/n the difference between the simple power law and the calculated flux 
is $\lsim$ 20\%. Below this energy, it is clear that the effects of energy changing, 
convection, halo size and spallations become more and more important in 
shaping the B/C ratio, as studied in Refs. \cite{usineI,usineIbis}.
\\
This figure wants to emphasize the discrepancy among the models 
- basically due to the $\delta$ values - 
with increasing energy. At energies $\sim$ 100 TeV/n the predictions for 
B/C ratio in the $\delta$=0.3 and $\delta$=0.85 case - both not excluded by low energy 
data, even if $\delta$=0.85 is a quite extreme value - 
differ by two orders of magnitude. 
Different predictions for $\delta$ lead to B/C whose relative values
increase  with increasing energy. This high energy region is thus very 
interesting from an experimental point of view, since it could bring  - at 
least in principle - to a clear determination of the actual diffusive regime.
\\ 
Finally, we want to underline that the injection spectrum has no relevance in the calculation 
of this ratio. We have checked that for E $\gsim$ 10 GeV/n the results for the B/C ratio
are practically unchanged if we vary $\alpha$ (equal for each nucleus)  
in the very large range 1.8-2.5, as well as if we fix $\alpha$ for each nucleus
($\alpha^j$, indeed) as derived in Ref. \cite{wsoo}.
\\ 
Following the procedure outlined in Sect. \ref{sim}, the expected number of carbon and boron
events can be derived by selecting a given set of experimental conditions (exposure, time
and resolution) and an input flux.
Fig.\ref{fi:BCattb} shows the B/C ratio as obtained from case 1 
(upper panel, corresponding to $\delta$=0.3) and from case 3 (bottom panel, $\delta$=0.6), 
for three different collecting powers. 
The maximum energy at which the B/C ratio can be measured 
with a significant number of events - we are requiring at least 10 events for Boron nuclei - goes 
from about 900 GeV/n to more than 10 TeV/n when the exposure increases from about 40 m$^2$sr 
days up to 10000 m$^2$sr days. The various correction factors which must be
considered in the analysis of real data (e.g. due to selection
efficiency or interaction losses in the apparatus) would have the effect of lowering the maximum detectable energy. 
As an example, a global efficiency of 30\% would shift it by a factor 1.5 for the
smallest exposure considered here and case 3 as input. 
The same effect is produced for balloon experiments due to spallation processes in the
residual atmospheric grammage of 4-5 g/cm$^2$.\\ 
\\
>From this figure it is evident that the two theoretical models could be easily distinguished already by
experiments with acceptance  of about 1 m$^2$ sr  and flying time of forty days  (in our example: 1.3
m$^2$ sr and 30 days data taking). Note however that the figure has been
obtained without considering the systematic uncertainties.
\\ 
The inclusion of a finite energy resolution of the detector in the simulation has important consequences
on the ability to measure the B/C ratio and to distinguish among different models: it produces in  fact
a distortion which reflects the change in the carbon and (more) in the boron spectra due to the energy
fluctuation. The steepness of the spectra makes it more likely to move  low energy events to high energy
bins, so that the effect is bigger for weaker spectra. In the case of model 1, for example, the measured
B/C will be some percent higher than what expected  for an ideal detector, that is a ``zero resolution''
one, almost constantly from 100 GeV/n on;  on the contrary, using the input spectrum of model 5 the B/C
ratio appears to be from 12 to 18\% higher when going  from 100 to 1000 GeV/n.  A correction for this
effect has to be applied before the comparison with expectations; the binning of the data should be
carefully chosen in order to be bigger than the uncertainty due to energy resolution.
\\
Using two different input models {\it j} and {\it k}, we simulated the expected number of carbon and boron 
events. The
difference between the B/C ratios so derived can be quantified by performing a relative $\chi^{2}$ test
\begin{equation}
\chi^{2} = \sum_{i}\frac{((B/C)_{ij}-(B/C)_{ik})^{2}}{\sigma_{ij}^{2}+\sigma_{ik}^{2}}
\end{equation}
where we put at denominator the variance 
of the difference. In order to fully exploit the high energy region which can be explored by the current and
future apparata, where only acceleration and diffusion dominate the cosmic
rays spectrum, the summation starts from 100 GeV/n.
\\
The ratio between the carbon and boron fluxes is only mildly sensitive 
to systematic errors (most of which are supposed to be independent from the considered 
nucleus), which are however to be included in the total error. Since a full discussion on systematics 
can only be done for each specific apparatus, we consider here only a constant systematic error to be 
added in quadrature to the statistical one.
\\
Fig.\ref{fi:probGF} shows the result for the data sets respectively expected from model 3 and each of the other
ones as a function of the systematic uncertainty, for three different experimental collecting powers.
 We should remind that even at the lowest energy of 100 GeV/n the
percentage difference between models is big, going from about 7\% for models 3 and 4 up to 30\% for
models 1 and 2. Exposures are such that even for the smallest experimental configuration considered here 
the statistical errors are quite small and in the ideal case of no systematic effects all models
could  be separated. 
As obvious, the sensitivity increases with  differences between the two compared $\delta$ values, in
our case the ones labeled (1-3) and (3-5) for which $(\delta_{j}-\delta_{k})$=0.30 and 0.25 respectively. For
example, a $\delta$=0.3 case can be distinguished from a $\delta$=0.6 case with 90\% C.L. if the
systematic error stays below 12\% even in the  less favourable experimental conditions considered.
In the case of a satellite-like exposure, these two models could be distinguished even in presence
of a 15\% systematic error. On the other hand, the range of sensitivity raises up to more than 20
TeV/n for satellite-based experiments, thus allowing to tag bench marks with $\Delta \delta$=0.10-0.15
(e.g. case 2 from 3 or case 3 from 4), but only if  systematic uncertainties are kept
below 10\%.

The analysis shown in Fig.3  is meant to illustrate the ability of a
given experimental setup to discriminate between two fixed and arbitrarily
chosen $\delta$ values. In order to evaluate the accuracy in the determination of
$\delta$, we now perform a $\chi^2$ minimization procedure on the simulated 
B/C data, leaving $\delta$, $K_0$,  $L$ and $V_c$ as free parameters.
As expected, the resulting $\chi^2$ has no structure with respect to  $K_0$,  $L$ and $V_c$, whose effect
is in practice reabsorbed in a global normalization of the  spectra.  On the contrary, the calculated
$\chi^2$s are such that a minimum for  $\delta$ can always be  clearly
identified.\\
If systematics can be considered negligible, the determination of the
diffusion  coefficient slope can be obtained at 10-15\% significance from B/C
data collected with the 39 m$^2$ sr days simulated experiment, and at 8\%-10\% for the 500 m$^2$ sr
days case. 
However, we find that $\delta$ can be determined with significance level of 15\% or
10\% respectively for the 500 m$^2$ sr days or 10000 m$^2$ sr days
simulated experiment, if the systematic errors are of order 10\%. Very low
accuracy is predicted in this case from the lowest exposure experiment (39
m$^2$ sr days).\\
As expected, the obtained accuracies are consistent with the probability curves shown
in Fig.3; e.g., the case 3-4 presented in this figure (bottom left) considers
two $\delta$ differing by 15-20\% and shows that they are perfectly
separated only if the systematic uncertainties lay below 2-3\% for the smallest
collecting power. This agrees with our previous conclusion, namely that the
accuracy on $\delta$ is at the level of 10\% for a collecting power of 
39 m$^2$ sr day and negligible systematics.

We emphasize once again that the above results are derived including systematic
uncertainties which are assumed to be independent on energy: while this is true for trigger efficiency 
or event reconstruction, or even for Monte Carlo corrections, other systematic errors like those due to
energy resolution will depend on energy and could alter the considered fluxes. 
\\
A spectral deformation of the B/C ratio at high energies has been proposed in
Ref. \cite{berezhko} as due to distributed reacceleration (see above). CR
particles can have sporadic encounters with SNRs, during which they could gain small
amount of energies.  The modification of the primary spectrum comes out to be
independent of energy, while the effect for secondaries increases with energy
just as the diffusion coefficient in the Galaxy. Thus, the ratio of reaccelerated
B/C$^{re}$ to standard B/C  would be roughly  described by 
(B/C)$^{re}$/ (B/C) $\propto (\alpha+\delta-1)/(\alpha+2\delta-1) (p/p_{GCR})^\delta$, 
where $p_{GCR}$ is the
momentum corresponding to the maximum of the observed particle spectrum  (in
Ref. \cite{berezhko} it is  assumed $E_{GCR}$=0.6 GeV/n). The real
situation is rather more complicated than this description, which nevertheless
gives us the correct trend and the possible magnitude of the effect. 
The expected deformation of the B/C ratio would be greater
for greater $\delta$. Moreover, as figured out in Ref. \cite{berezhko},
 its intensity would strongly depend on the value of the
circumstellar hydrogen density, $N_H$. The higher $N_H$, the lower the
distortion. In the case of $\delta$=0.6, 
an effect of one order of magnitude would be expected at 1 TeV/n for
$N_H$=0.003 cm$^{-3}$, a very low number, indeed. 
For higher and more plausible values of the
hydrogen density, the B/C ratio could be enhanced by a factor of two.
If the diffusion is Kolmogorov-like, only very small $N_H$ could give an observable
effect (about a factor two). But in this case it would be less ambiguous to 
correctly interpret data, while in case of higher $\delta$ this sporadic reacceleration
effect could be mimicked by smaller $\delta$ effects. 
\\
Non-stationary models have also been studied, taking into account the fact that SN 
might be considered discrete sources  in the Galaxy, so that a time-dependent diffusion equation 
should be solved \cite{riri}. However, it seems plausible that the B/C ration would not be modified 
in shape, but only in a overall normalization factor \cite{bus}.

\subsection{Primary fluxes and acceleration spectra}

Primary fluxes trace back to the diffusion and acceleration properties, since they are
produced directly in the acceleration sites. 
Charged CR nuclei heavier than protons suffer catastrophic losses due to
nuclear destruction on the interstellar H and He. These reactions become irrelevant on the
CR spectrum  at energies of about 1 TeV/n, depending on the nuclear species. 
The effect of inelastic interactions  lowers and flattens the flux, giving rise to a
final shape that is harder than a pure power-law $E^{-\gamma}$, being $\gamma=\alpha +
\delta$. 
\begin{figure*}[!htbp]
 \includegraphics[viewport=10 50 550 400,width=.99\textwidth,clip]{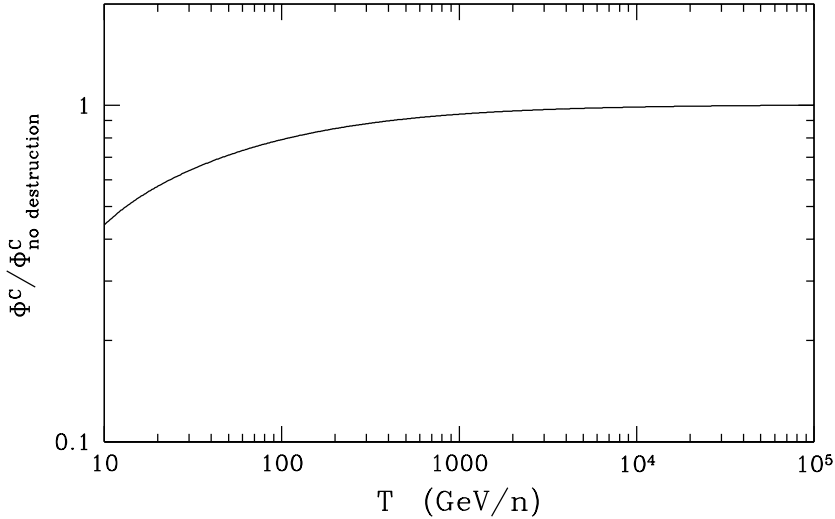}
 \caption{\it{Ratio of calculated carbon fluxes from the same propagation model, where nuclear destructions 
have been turned off at the denominator.}}
\label{fi:C_ratio}
\end{figure*}
This effect has been studied by computing the carbon flux according to the full diffusion model 
(case 2 and $\alpha=2.05$, but in this context their value is not relevant) 
with and without inelastic destructions; the result is plotted in Fig.\ref{fi:C_ratio} 
(for iron flux, see Fig. 1 in \cite{david_HE}).
The difference between the two cases is about 30\% at 100 GeV/n, and at 1 TeV/n it is still 5\%. 
Thus, it is in principle not correct to expect a pure power law spectrum from energies 
above few tens of GeV/n, as often claimed in literature. Of course, the error bars on the existing
experimental data for primaries (except for protons, maybe) still justify this approximation. \\
>From the experimental point of view, data about fluxes of primary nuclei were collected in the past
for Z$\leq$26 up to few TeV/n, but with decreasing statistical significance as the energy
grows higher.
At the highest energies, the only available data sets come from emulsion chamber
experiments \cite{jac,runj,Sok} which measure groups of elements; systematics are quite
large, e.g. the results for the CNO group differ between JACEE and RUNJOB by a factor 
of 2 in normalization. More recently, the first data from a one day test flight of TRACER have been 
published for Z$\geq$8 nuclei \cite{trace}.
\\
We have calculated the carbon flux for a propagation model in agreement with 
the B/C analysis presented above, with the only aim to qualitatively reproduce 
the data at the energies in discussion. So we do not care much about the level of 
 agreement with lower energy data, aware that a thorough study of the propagation 
of primaries at low energies would require many deeper insights, which are 
beyond the scope of the present paper. 
Figure \ref{fi:Cflux} shows a compilation of the data on carbon flux together with the
calculated flux for model 3 ($\delta$=0.6) and $\alpha$=1.9, 2.05 and 2.2.
The normalization has an important role 
in this calculations and  may be ascribed to the value of the source abundances, which are 
very poorly known, and of $K_0$. We refer our normalization to the fluxes measured at 35 GeV/n \cite{Heao},
where the total error (statistical and systematic) is about 15\%. \\
\begin{figure}[!htbp]
 \includegraphics[width=0.99\textwidth]{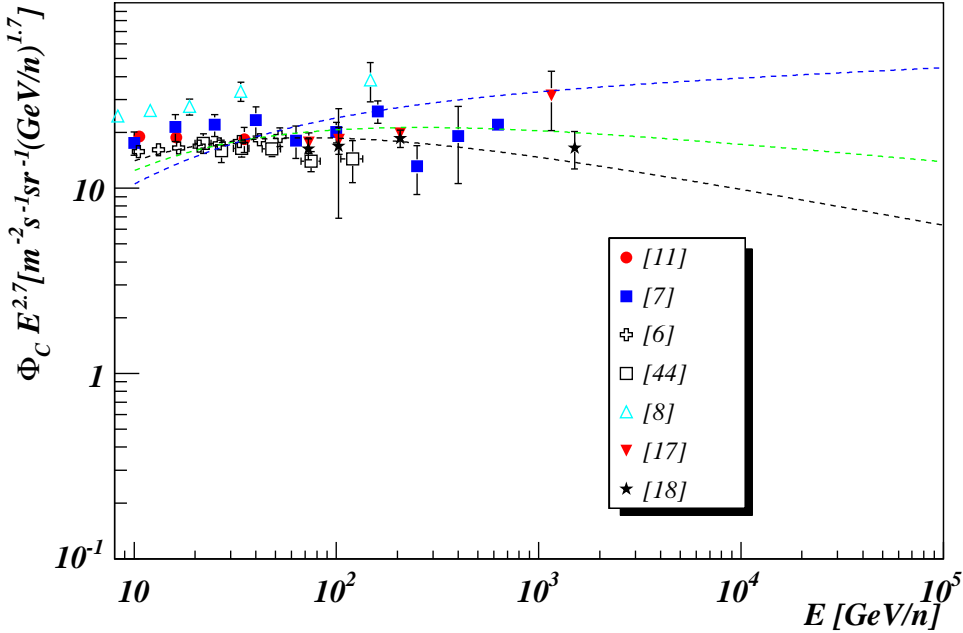}
 \caption{\it{Data on carbon flux multiplied by E$^{2.7}$; 3 theoretical 
models are shown, corresponding to  $\delta$=0.6 and $\alpha$=1.9,2.05,2.2 from top to
bottom, respectively.}}
\label{fi:Cflux}
\end{figure}
\\
The expected number of events from the above described model and for different values of
$\alpha$ have been calculated following the procedure described in Sect.\ref{sim} for different
experimental exposures; the corresponding flux is shown in
Fig. \ref{fi:Catta}. 
\begin{figure}[htp]
\centering
\includegraphics*[width=0.99\textwidth]{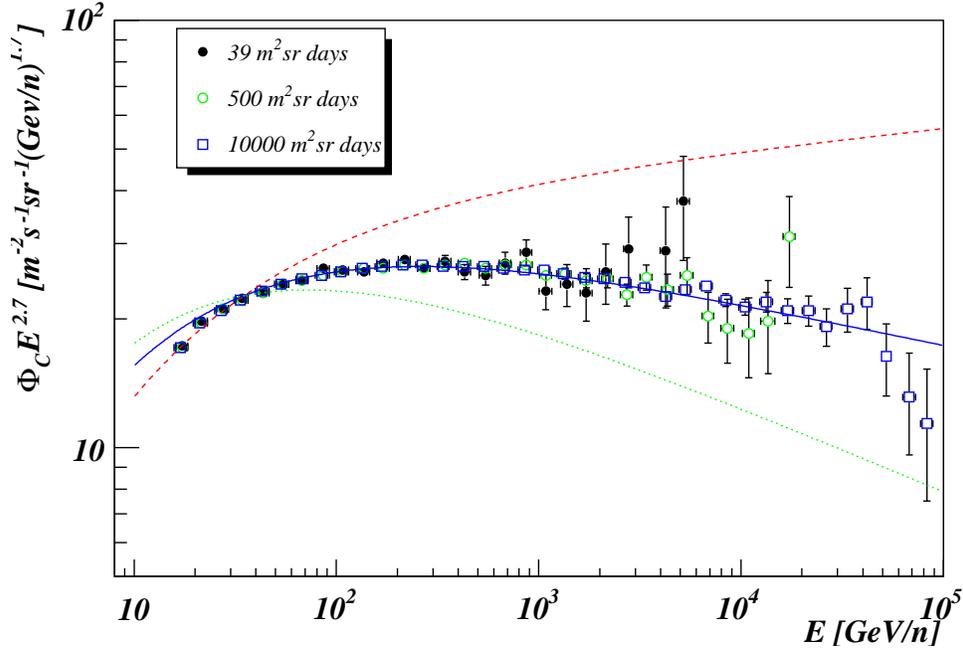}
\hfill
\caption{\it{Expected C flux for input model with $\alpha=2.05$ and $\delta=0.6$ and 3
different exposures. Theoretical curves as in Fig. \ref{fi:Cflux}.}}
\label{fi:Catta}
\end{figure}

We have also checked that this flux is almost independent of the assumed energy
resolution by changing it from a constant value to an either decreasing or increasing behavior
as a function of the particle energy (from 40\% at 100 GeV/n to either 50\% or 30\% at 1000 GeV/n).
Indeed, it can be shown that the energy resolution plays a minor role as compared to the event statistics 
(to collecting power) in the estimation of spectral slopes for  simple power laws \cite{how}.
\\
The statistical error on the expected events is less than 2\% up to about 500 GeV/n, rising to
10\% at 2 TeV/n for the smallest exposure, with 500 m$^2$sr days it lowers to few percent at 
1-2 TeV/n: the new detectors will so have full ability to cover the high energy range with 
sufficient statistical precision.
With a $\chi^{2}$ minimization procedure, the expected data are fitted to a power law and 
the slope $\gamma$ is determined with an error lower than 1\% in a range $\Delta$E=0.1-10 TeV/n. 
The effect of adding a constant systematic error of 10\% is of more than doubling the error on $\gamma$.
\\
This result implies that the experimental errors on the carbon flux (the same conclusion is reached for
oxygen) will not add any uncertainty in the determination of the
acceleration power law index $\alpha$  other than the one carried by the diffusion
slope $\delta$. If the latter is determined within 10\% from B/C
measurements, the same experiment will let us fix the former within an equal 
uncertainty level by means of primary fluxes determination. 
\\
The knowledge of $\alpha$ will lead to a better understanding of the acceleration processes inside 
galactic supernova remnants. In particular, we note that even in a LDB mission
it could be possible to reach 1 TeV/n with quite good statistical accuracy, thus allowing to
confirm or exclude whether $\alpha$ lies towards low-values 1.9-2.1, or in the higher range 2.3-2.4
at a 3 $\sigma$ level.

\section{Conclusions}
\label{concludiamo}

A two-zone diffusion model for stable, galactic CRs  has been extensively used to generate 
different predictions on the primary and secondary nuclei fluxes, which in turn are used to
evaluate the expected events in various experimental configurations similar to present or next
generation apparata. They will substantially exploit a high collecting power in the  high energy
region ($\geq$ 100 GeV/n), where the low energy effects and most of the peculiarities of the model
can be neglected. We have shown that new
detectors exploiting the long and ultra long duration balloon flights, or even more future space
missions, will provide valuable information about both  the diffusion coefficient slope $\delta$  and the
acceleration spectrum $\alpha$ of galactic CRs. 
\\
An experimental determination of  $\delta$ through the
measurement of the B/C ratio is at reach of the new generation of detectors, some of which
are already taking data. An experiment like CREAM - which during its first flight collected data for
more than 40 days breaking all the previous flight and duration 
records\footnote{http://cosmicray.umd.edu/cream/cream.html} -  is in  principle able to distinguish
between a Kolmogorov-like or a different turbulent regime, determining $\delta$ with an uncertainty
of about 10-15\%, if systematic errors are low enough. If the latter were 
important and at a level of about 10\%, the uncertainties on the derived $\delta$ would be huge.
Experiments reaching collecting powers of hundreds  m$^2$ sr days have the potential of fixing $\delta$ 
at about 85\% confidence level even in the case of important systematic errors, or at $\gsim$ 90\% 
if the latter are of the order of the statistical ones. 
\\
Only  space-based  or satellite detectors  will be able to determine $\delta$ with very high
accuracy. In fact, even if they would be affected by not negligible systematic errors, the covered 
energy range would be significantly wider (we discussed the case with maximal energy of order 30
TeV/n).
\\  
A careful measurement of the single fluxes will shed new light on the
acceleration spectrum: in the high energy region the new generation apparata will be able to
measure the spectrum slope $\gamma$ for light nuclei with a negligible error. 
No further uncertainty other than the
one affecting $\delta$ as derived from the B/C ratio will then be added to the 
determination of the acceleration slope $\alpha$.  This last parameter is thus expected to be
measured with an uncertainty of 10-15\% in the near future, allowing to exclude large ranges of
$\alpha$ values and  leading to a much clearer understanding of the acceleration processes.
\\
The spectra of the single elements of the CR beam will be measured with  very good
statistics up to more than some 10$^{14}$ eV (which is the correct unit to be used when
comparing with indirect measurements, for which only the energy/particle is determined). 
With a careful analysis of the
experimental systematics, absolute particle fluxes will be provided, thus giving more than a
decade calibration region to the Extensive Air Shower detectors.  
Direct measurements based on adequate statistics in the region of overlap with the Extensive Air
Shower explored region are of main importance in order to give ground based experiments a firm
reference point, thus  helping in deriving the mean composition of the CR beam towards
and at the knee and in  checking the models which are used to derive the energy.  Around
10$^{14}$ eV, the flux uncertainty can  be set at present around 30\%; with the new
experiments now under way or in project, we can expect to  significantly reduce it.
The same is true for what regards the mean mass: results from indirect measurements are far
from clear \cite{ahr}, while data from direct detection still show about 30\% uncertainty
above 10$^{14}$ eV, which can surely be lowered at least to 15-20\% with the new apparata. 
\\
At high energy, above 100 GeV/n,
the primary nucleons contribute mainly to neutrino-induced upward through-going or stopping
muons \cite{SK}; most important are protons and helium nuclei, while  heavier ones contribute
less than $\simeq$10\% at all energies. Apparata with the considered exposures will be able 
to measure also proton and helium spectra with unprecedented accuracy, thus allowing to reduce
the present uncertainty  in the all nucleon spectrum, which including all the available data is
still around 30\% in the TeV range. 
\\
We want to emphasize that a better understanding of the propagation properties is of great significance
for the study of signatures of new physics in CRs. Indirect signals of dark matter pairs annihilating in
the halo of our Galaxy could be found  in antiproton, antideuteron or positron CRs. This research is
limited by the uncertainties in the propagation parameters, that dramatically affect the fluxes of
charged particles located in the whole diffusive halo \cite{pbar_susy}. For instance, if this
uncertainty  were significantly reduced, it would be possible to exclude supersymmetric models
predicting antiproton fluxes exceeding the present measurements \cite{leggeri}. A better
signal--background discrimination would come  from antideuteron measurements \cite{dbar}, and a
reduction of the astrophysical uncertainties on the calculated fluxes would be even more desirable. 
\\
We conclude  by reminding that  important parameters -- other than $\delta$ and $\alpha$ --  describing
our Galaxy in relation to the propagation of charged particles can be fixed only with very precise low
energies data, and for various species. This in principle could be done by the AMS experiment on the
Space Station \cite{ams}, which could measure fluxes from hundreds MeV/n up to TeV/n and for 1$\lsim$ Z
$ \lsim$ 26.

\section{Acknowledgements} 
We warmly thank D. Maurin and R. Taillet for the usage of numerical code for cosmic  ray propagation
developed in collaboration with one of the author (F.D.), and  M.Aglietta, B.Alessandro, T. Janka, P.S.
Marrocchesi, D. Maurin and G. Raffelt for useful discussions.  F.D. acknowledges the A. von Humboldt
Stiftung for financial support while at MPI in M\"unchen.

\end{document}